# Theory-driven automated content analysis of suicidal tweets

## : Using typicality-based classification for LDA dataset

Joon-Mo Park[1], Chul-joo Lee[1], Yunseok Jang[2]

[1] Department of Communication, Seoul National University

[2] Department of Computer Science and Engineering, Seoul National University

**Abstract**

This study provides a methodological framework to classify tweets according to variables of the Theory of Planned Behavior. We present a sequential process of automated text analysis which combined supervised approach and unsupervised approach in order to detect one of TPB variables in each tweet. We conducted Latent Dirichlet Allocation(LDA), Nearest Neighbor, and then assessed "typicality" of newly labeled tweets in order to predict classification boundary. Furthermore, this study reports findings from a content analysis of suicide-related tweets which identify traits of information environment in Twitter. Consistent with extant literature about suicide coverage, the findings demonstrate that tweets often contain information which prompt perceived behavior control of committing suicide, while rarely provided deterring information on suicide. We conclude by highlighting implications for methodological advances and empirical theory studies.

**Introduction**

Every year an estimated 788,000 people kill themselves worldwide and in South Korea alone more than 10,000 people commit suicide annually (KOSTAT, 2015; WHO, 2015). Many studies suggest that the way media depicts suicide can have an influence on people's attitude towards suicidal behaviors, which is often represented as suicide contagion or "Werther effect" (Philips, 1974). This risk is thought to depend not only on victim's personal characteristics but also on the volume of coverage and description of suicidal behaviors (Tatum, Canetto & Slater, 2010).

To date, the greater part of evidence for suicide contagion is found in studies on traditional media (Lee et al., 2014; Romer, Jamieson & Jamieson, 2006). However, recent public health studies take into account not only traditional media but also social media as an influential source of suicide contagion (Luxton, June & Fairall, 2012). In Korea, suicide is the leading cause of death among teens to thirties (KOSTAT, 2015) and there are several ways social media can increase risk of suicide, especially to young people. One possible reason for suicide among young people is, large quantities of suicide-related information shared via social media. Considering that 90% of young adults use social media (Perrin, 2015) and they are psychologically more vulnerable with higher risk behaviors (Dobson, 1999), the volume of suicide description and suicidal information on lethal means to kill oneself are highly likely to have an impact on young adults. The other is disinhibition that people reduce preexisting restraints on the specific behavior by watching others commit suicide in media (Romer, Jamieson & Jamieson, 2006). Since a large number of messages including suicidal experiences are spread through social media, it is likely to lead people to a higher risk of suicide by reducing doubts or fears on committing suicide.

To deepen our understanding of the potential impact of social media on suicidal behaviors, we performed an automated content analysis on suicide-related tweets, grounded on the theory of planned behavior (TPB) (Ajzen, 1991). According to the TPB, individual beliefs on certain behavior predict intention to perform that behavior. To identify potentially encouraging or deterring messages that can affect individuals' beliefs on suicide, we classified suicidal tweets according to TPB variables such as attitude, subjective norm and perceived behavior control(PBC).

The specific aims of this article are twofold. First, we seek to analyze what kinds of suicidal messages are distributed on Twitter. Some studies attempted to analyze linguistic features of victims who committed suicide (Gunn & Lester, 2015; Stirmen & Pennenbaker, 2001), while others examined suicide coverage in traditional media (Gould et al., 2014; Schäfer & Quiring, 2015; Tatum, Canetto & Slater, 2010). However, few studies have applied behavior change theories to assess possible effect of suicidal messages. Thus, we look forward to inferring the potential effect of Twitter usage by revealing theory-based components of suicide-related tweets. The second is to develop an automated way to classify large amounts of tweets with minimal amounts of human-annotated data. Although computational text analyses help to scale up the amount of corpus by reducing calculation costs, researchers often face several impediments when attempting to capture latent meanings in complex semantic structures such as metaphors or sarcastic expressions (Shutova et al., 2017). To deal with this problem, many researchers have applied supervised learning algorithms, in which computer learns linguistic patterns from manually annotated documents to classify unlabeled documents. In this process, a large quantity of manually annotated documents is required in order to learn sufficient information for which to classify. This makes supervised learning approach dependent on the coverage or availability of data resources, limiting the application of supervised learning in the context of behavior change theories. The lack of annotated corpora for the computer to classify text dataset according to behavior change theories makes this approach difficult to apply. Furthermore, fewer lexical resources are available which can be applied to suicidal issue and even fewer in another language such as Korean. As a solution to that problem, we conducted subsequent automated analyses with a small number of human-annotated documents. In this study, we took steps toward developing automated text analytic models for detecting TPB variables in tweets, using LDA (Blei, Ng, & Jordan, 2003) followed by the Nearest Neighbor (Cover & Hart, 1967). Then, researchers engaged to judge the scope of clusters with typicality measurement.

## The role of theory in content analysis

Content analysis is a research method that aims to understand patterns of messages (Krippendorff, 2012; Manganello & Fishbein, 2008). Even though content analysis does not necessarily require theoretical background, the use of theory can be beneficial for two reasons. First of all, it is more efficient because theory offers guidance for selecting both constructs and methodological approaches. In its absence, research designs, variables, measurements and hypotheses are likely to be too arbitrary and fragmented to cumulate in a meaningful way (Steinfield & Fulk, 1987). The second benefit is, theory can provide a basis for combining content analytic findings with media effect studies (Manganello & Blake, 2010; Manganello & Fishbein, 2008). The ability to propose explanations is an important factor in media studies, and theory-based content analysis is likely to make contributions to further research by describing information environment we are exposed to (Steinfield & Fulk, 1987). Content analysis itself provides description of media content and assesses the amount of a particular media type, however, once theory-based variables of media content are identified, then researchers can combine content analytic data with survey data to build an argument for media effects (De Vreese et al., 2017).

## The Theory of planned behavior

TPB is applied to this content analysis to investigate the prevalence of persuasive appeals in suicide-related contents. This theory predicts behavior intentions in many different areas such as opinion expression (Neuwirth & Frederick, 2004), anti-smoking (Cohen, Shumate & Gold, 2007) or intention to register as organ donors (Bresnahan et al., 2007). TPB states that an individual's intention to conduct a specific behavior predicts his or her actual performance of the behavior (Ajzen, 1991). Intention is, in turn, a function of three determinants: attitude toward performing the behavior, subjective norm, and PBC (Ajzen, 1991; Fishbein & Cappella, 2006).

Attitudes, subjective norms, and PBC themselves are assumed to be based on underlying beliefs (Fishbein & Cappella, 2006). Attitudes are functions of beliefs about whether performing the behavior will lead to good or bad outcomes. For example, the more one believes that performing the behavior will lead to positive consequences, the more favorable will be the person's attitude. Subjective norm is a function of beliefs that significant others think one should (or should not) perform the behavior as well as others in his or her social networks are performing (or not performing) the behavior (Cohen, Shumate & Gold, 2007). The more one believes that others think one should perform the behavior and that behavior is perceived as prevalent to them, the stronger will be the subjective norm to perform that behavior. PBC is a function of beliefs that one can perform the behavior even in the face of specific impediments (Ajzen, 1991). Therefore, messages depicting an individual who can carry out such behavior in the face of barriers may increase perceived behavior control, leading to a higher-level behavior intention.

In general, TPB aims at explaining cognitive mechanisms of normal behaviors (Hales, Householder, & Greene, 2002). Even though suicidal behavior has long been considered as "abnormal" which is merely caused by mental dysfunction, it is revealed that suicidal behavior is not just prompted by irrational impulse of the moment (Van Heeringen, 2001). Rather, suicide is explained in terms of normal psychological constructs since it involves large amounts of cognitive processes such as searching for the method, estimating possible influence on others, and comparing merits and demerits of committing suicide (O'conner & Armitage, 2003). In this regard, committing suicide is within the range of the theory's application which is carried out at the end of conscious decision making (O'conner & Armitage, 2003).

## Computational content analysis in the context of suicide

Compared to analyzing textual documents written in formal language (e.g. news article), analyzing suicide-related tweets poses unique challenges; tweet is relatively short (140 characters or less) with language unlike standard words on which many supervised learning models have trained and evaluated (Ramage, Dumais & Liebling, 2010). In addition, frequent metaphorical expressions in suicide-related dialogues represent a significant challenge for conducting dictionary-based text analysis (Reeves et al., 2004; Shutova et al., 2017).

In general, there are three different approaches for automated text analysis: manual dictionary, supervised learning and unsupervised learning (Grimmer & Stewart, 2013). For many years, automated analyses of suicidal contents have developed around the use of manual-dictionary (Stirmen & Pennenbaker, 2001). This approach counts the number of words related to the concept as defined in a dictionary, then calculates a score for each category (Grimmer & Stewart, 2013). For example, Stirmen and Pennenbaker (2001) looked through works of poets who killed themselves, and analyzed whether their poems included more words with respect to themselves and fewer words pertaining to the collective.

Dictionary approach has been regarded as the way to conduct content coding corresponding to theoretical variables since researchers can set categories on the basis of theoretical expectations (Schwartz & Ungar, 2015). However, this method falls short of analyzing complex semantic structures such as metaphor or sarcasm. Discerning metaphoric or sarcastic messages is very difficult for machines because it requires knowledge of the topic and even the users themselves (Nobata et al., 2016). For instance, "Suicide is like wrapping your pain as a gift and hand them to the loved ones"– which is not literally true but means that suicide is painful to his or her loved ones – is encoded as suicide referring to "gift". Even though metaphoric expressions can be interpreted in a totally different way under the specific domain, dictionary method cannot deal with these word usages because it relies on surface meaning of each word.

Supervised learning also enables researchers to classify text data according to predetermined categories. A computational algorithm learns from a set of annotated documents which is called the train set, and then the classifier is used to classify documents in the test set. The key for supervised classification is extracting features from train set which are indicative of each label. The larger the train set, the more features supervised learning algorithm learns with which to classify test set. However, applying supervised approach to classifying millions of unlabeled tweets is costly as it requires a large, almost prohibitive, number of human-annotated examples to learn accurately (Nigam, McCallum, Thrun & Mitchell, 1998; Petchler & González-Bailón, 2015).

Unsupervised learning inductively identifies patterns from unlabeled data by clustering documents that contain same words in common (Burscher et al.,2014; Guo et al., 2016). Unsupervised learning methods are often used for exploratory purposes since they can identify patterns of text that may be theoretically useful but unknown to researchers (Grimmer & Stewart, 2013). In communication studies, topic modeling stands at the forefront of unsupervised learning methods. Topic modeling assumes that documents sharing similar topics are likely to use a group of similar words (Blei, Ng & Jordan, 2003). Latent topics can be detected by identifying a group of words that frequently occur together within each document. However, the result might not necessarily correspond to the theoretical categories, since it solely derives topics from stochastic models. This can restrain the researcher from classifying text dataset according to the elements of a specific theory. Latent Dirichlet Allocation (LDA) is one of the most popular topic modeling techniques for unsupervised learning. Many prior studies conducted LDA to extract features of the data and learn information about the semantic structure. For instance, Guo and colleagues (2016) conducted LDA on 7.7million tweets mentioning "Obama" and "Romney". They found that the LDA performed better than the dictionary-approach especially on identifying more nuanced meanings of the message. However, as some LDA topics included multiple issues in one topic, they claimed that human evaluation would be helpful to validate LDA results.

(Figure 1)

Given these advantages and disadvantages of each method, this study combined supervised learning and unsupervised learning in order to take advantages of both approaches. The overall process is illustrated in Figure 1. Supervised learning (e.g. 1-NN) enables researchers to set categories which represent variables of specific theories, while unsupervised learning (e.g. LDA) requires less human effort to conduct and extract features of each topic. Through combined automated text analysis, we aim to analyze suicidal tweets including information on attitudes, subjective norms and PBC that can further influence individual suicidal intention. As a result, the research questions are as follows:

RQ1: What kind of TPB variables are most prevalent in suicide tweets?

RQ2: How can the computer detect TPB variables in each tweet?

RQ3: Which variables are accurately detected in our model?

## Method

**Data Collection**

Suicidal tweets were obtained via Twitter REST API with the keyword "suicide" in Korean. Twitter provides access to data through three different API's: REST, Search and Streaming (Sinhura & Sandeep, 2015). Developers have access to data which includes tweets, status data, user information, and timelines by using Twitter REST APIs. The collection started on August 19, 2016, and ended on September 23, 2016. In all, approximately 3.1 million tweets were retrieved. As preprocessing steps, we cleaned datasets by removing irrelevant information or replacing them into standardized forms. We removed 'RT' and timestamp, while hyperlinks were replaced into 'URL' and tweet mentions into 'MENTION'. Telephone numbers were converted into 'NUMBER'. After removing unrelated information, we removed tweets shorter than five words since overly short sentences cannot fully express specific meanings (Lee et al., 2017). The filtered dataset contained 1.4 million tweets.

**Categories**

The body of suicide-relevant tweets was coded into the TPB variables (attitude, subjective norm, PBC). Each tweet was coded as representing one of thirteen variables: *positive outcome, negative outcome, approval of significant others, disapproval of significant others, approval of others, disapproval of others, descriptive norms, ease of suicide, difficulty of suicide, mention of specific methods, mention of specific place of suicide, sources of help against suicide, sources that promote suicide*.

*Attitude*

The first persuasive suicide message appears in it's attempt to influence the individual's opinion on the behavior. This type of tweet emphasizes characteristics of the desirable behavior. We provide two types of tweets coded in this category; *positive outcome* and *negative outcome.* Tweets referring *positive outcome* include depiction of suicide as a solution to a problem, eternal rest, an escape route, the only option left or a pain reliever. *Negative outcome* tweets contain information on falling into hell, imprisonment or abuse of corpse as a result of suicide.

### Subjective Norm

The second types of tweets under consideration are messages indicating social pressure. If the tweet depicts either approval or disapproval of others in the society, it is classified as altering injunctive norm. Injunctive norm was coded in four different ways as *approval/disapproval of significant others* and *approval/disapproval of others*. Significant others refer to victim's parent, guardian, sibling/cousin, friends/peers, teacher, partner, partner, health provider, or religious leader. Descriptive norm motivates others by informing individuals of prevalent action in a situation. If tweets contain information on how often people commit suicide or mention celebrities who committed suicide, they are identified as *descriptive norms*.

### Perceived Behavior Control

The third persuasive component is PBC which have an impact on individual's belief that he or she can accomplish suicidal act. Variables with regard to PBC are *ease of suicide, difficulty of suicide, mention of specific methods, mention of specific place, sources of help against suicide, sources that promote suicide*. If the tweet discusses ease, feasibility or the low cost of committing the suicide act, it is annotated as *ease of suicide* whereas depicting hardships or failures in committing suicide is annotated as *difficulty of suicide*. If the content informs the readers of a specific method to kill themselves, it is annotated as *mention of specific methods of suicide* (e.g. shooting oneself, jumping from height, hanging, suicide bombing, etc). In addition, tweets informing the readers of a specific place (e.g. river, bridge, building, railway, etc) to commit a suicide are annotated as *mention of specific place of suicide*. If the tweet contains information about specific source of help (e.g. suicide prevention program, life line number, link of suicide-preventive website, etc) that will inhibit committing suicide, it is annotated as *sources of help against suicide*, while tweets inform the specific route that will promote suicide act (e.g. link of pro-suicide website, address of mass suicide clubs) are annotated as *sources that promote suicide*.

(Table 1)

In order to construct a train set, we randomly sampled 100,000 distinct tweets out of the whole dataset. Among 100,000 tweets, a total of 3,530 tweets were manually coded as one of TPB variables.[1] During this process, to assure the exclusivity of the sample tweets, human coders labeled tweets as one of TPB variables even though some tweets include diverse opinions on suicide. The samples of tweets and distribution of train set ($n = 3,530$) are provided in Table 1.

**The first step: Conducting latent dirichlet allocation**

We first built on LDA topics as proxies for TPB variables. A Python package "Gensim" (Řehůřek & Sojka, 2010) was used to train LDA. In our study, we identified the latent topics and words referring to each topic using the LDA with Gibbs sampling (Blei, Ng, & Jordan, 2003). In the hope that some topics correspond with TPB variables, we decided the number of topics as 100, which generally produces coherent topics (González-Bailón, S. & Paltoglou, G., 2015). The LDA identified a list of 100 topics and calculated probabilities of the words that are assigned to each topic. To determine whether there were topics which represent TPB variables, the researcher read all the corresponding words whose probability was higher than 1% and suggested a label that represented the TPB variables. Table 2 is the typical words of each topic extracted from LDA.

---

[1] A set of 90 suicide-relevant articles were double-coded to establish codebook reliability (Huh, 2017). Inter-coder agreement for TPB variables was assessed using percent agreement and Krippendorff's alpha (Krippendorff, 2012). The percent agreement on thirteen variables ranged from 95.40% to 100%, and Krippendorff's alpha ranged from .81 to 1.0.

(Table 2)

**The second step: Using nearest neighbor to annotate unlabeled tweets**

After distributing labeled tweets and unlabeled tweets together in shared semantic space through LDA, we generatively categorized unlabeled tweets into the same category with the closest labeled tweet. Nearest Neighbor (1-NN) was conducted in order to classify tweets into one of TPB categories, using manually annotated tweets to guide the learning process. The KNN algorithm classifies objects based on closest training examples, thus it can be beneficial when there is little knowledge about the distribution of the data (Domeniconi, Peng & Gunopulos, 2002). However, the performance of the K-NN classifier is largely influenced by the neighborhood size K. If K value is1, which refers to Nearest Neighbor, the estimate is likely to be poor because of the sparse distribution of data or mislabeled training set. Larger K value may deal with that problem, however, an increased K value is likely to degrade the classification performance owing to the inclusion of the outliers from other topics. To deal with this shortcoming, some studies tried to improve the K-NN performance by "typicality" judgment (Zhang, 1992).

**The third step: Judging whether newly annotated tweets have similar patterns with human-annotated tweets**

Previous studies (Bappy et al., 2017; Caddigan, Choo, Fei-Fei & Beck, 2017) revealed that the "representativeness" or "typicality" of an annotated data predicts the likelihood that the judgment will be accurate, as well as reducing the annotation cost. The concept of "typicality" originates from a psychological literature on categorization (Rosch, Simpson & Miller, 1976), which refers to the degree of the object to be judged as representative examples of specific category. Joffe and Haarhoff (2002) applied typicality to study Ebola-related themes pervaded in newspapers and interviews, arguing that "the typicality of a theme, even in a non-representative sample, provides an indication of the degree to which it is shared in the sample" (p. 959). Even though text analysis based on typicality is uncommon in communication studies, many computer vision studies utilized typicality value in order to detect objects in visual scenes (Fei-Fei & Li, 2010; Maxfield, Stalder & Zelinsky, 2014).

(Table 3)

Applying "typicality" concept to our research, human-annotated tweets functioned as typical (or representative) tweets of each TPB element. According to Maxfield, Stalder and Zelinsky (2014), typicality can be calculated with similarity distance from human-annotated objects to unlabeled objects, and this value can predict classification boundary. In this study, typicality of each tweet was predicted by calculating its average similarity to the human-annotated tweets in each category. Table 3 shows how the judgment on each variable changes as typicality value alters from 0 to 1. The value 0 refers to a perfect match with word distribution of human-annotated tweets and unlabeled tweets, while a higher value represents a complete mismatch between typical tweets and unlabeled tweets. As a result of this procedure, we determined that a typicality threshold of 0.275 works well. The final dataset comprised of 214,570 tweets. Figure 2 illustrates the idea of typicality-based classification for all the data sets used in our analysis.

*Typicality measurement*

$$d = \sqrt{(|Y - X|^2)} \text{ for each unlabeled tweet} \quad (1)$$

Y : word proportion of unlabeled document Y

X : word proportion of labeled document X

(Figure 2)

**Validation**

We conducted manual coding to assess the accuracy of automated-classification. We respectively calculated the Krippendorff's alpha for thirteen variables of TPB. Although this metric has been mainly applied to measure the agreement among human coders, it still offers a benchmark for assessing accuracy of automated content analysis (Gonzlez-Bailon & Paltoglou, 2015). To calculate accuracy, 25% ($n = 54,730$) of the final dataset ($n = 214,570$) were randomly selected. Among the 54,730 tweets, two coders independently coded approximately 20% of the tweets and examined whether the coding rule is reliable to evaluate the complete sample of tweets.[2] Then 54,730 sample tweets were divided by half, and each coder evaluated whether the computational classification result was accurate or not.

## Results

**Topic proportions of Twitter**

---

[2] The percent agreement of TPB variables ranged from 86.69% to 98.90%, and Krippendorff's alpha ranged from .74 to .97.

214,570 tweets out of 1.4 million tweets were in the scope of TPB clusters. Therefore, the remaining tweets (1.2 million) represent data which are related to suicide but not including TPB variables. We first examined the extent to which TPB-based persuasive suicide tweets were present in Twitter. Among the 214,570 tweets that are detected as including one of TPB variables in the content, more than three quarters of the tweets (78.11%) contained information that would either directly or indirectly have an impact on reader's PBC (i.e. ease, difficulty, method, place, source of help, source that promote suicide). Specifically, tweets mentioning specific methods of suicide act was most frequent (39.89%), followed by tweets providing sources that promote suicide (16.59%) and portraying suicide as easy to take action (14.73%).

(Table 4)

Approximately about one tenth of tweets (9.81%) contained information that can change reader's perception of descriptive norm regarding suicide. In addition, about 6.06% of tweets included information that affects reader's injunctive social norm about suicide; in particular, 5.59% of tweets portrayed negative injunctive norm regarding suicide (significant others' disapproval, general others' disapproval) while mere 0.47% of tweets depicted positive injunctive norm. Six point zero two percent of tweets include information related to attitude toward suicide (positive outcomes, negative outcomes); to specify, only 0.26% of the tweets depicted negative outcomes of suicidal behaviors while 5.76% depicted positive outcomes.

**Accuracy of automated TPB categorization**

Overall accuracy rate of the classified tweets is 74.77%. Three elements referring to *subjective norm* or *perceived behavior control* have been categorized with higher than 80 percentage of accuracy: *approval of other*s, *descriptive norms* and *sources that promote* *suicide* (Table 4). These high accuracy rates can be attributed to the factors such as web address (sources that promote suicide), statistical figures (descriptive norm) and linguistic style connoting approval (approval of others), which are easily detected compared to other features.

**Discussion**

This research presents detailed descriptions of which persuasive elements are prevalent in suicidal tweets. We used TPB variables to examine what kind of persuasive elements are likely to influence on people's suicidal intention. In order to classify large-scale text data into thirteen categories with a small number of annotated data, we combined two different computational learning approaches: supervised learning and unsupervised learning. Be worthy of notice is applying "typicality" concept which regard human-annotated tweets as the most typical tweet of each TPB variable. We assessed whether classification results accord with human judgment by calculating similarity between human-annotated tweets and unlabeled tweets.

Our study revealed that tweets often detailed suicide methods (39.89%), sources that promote suicide (16.59%), and portrayed suicide as easy to take action (14.73%). It provides empirical evidence that Twitter is used to disseminate information on how to commit suicide. Moreover, links to pro-suicide websites are widely shared on Twitter that can encourage vulnerable individuals to join the extreme community (Luxton et al., 2012). Meanwhile, it rarely provided deterring information on suicide: negative outcomes (0.26%), difficulty of suicidal behavior (2.37%), significant others' disapproval of suicide (1.86%) and general others' disapproval of suicide (3.73%). These results align with previous studies (Gould et al., 2014; Tatum, Canetto, & Slater, 2010) demonstrating that suicide coverage often includes information on lethal methods to kill themselves but rarely mentions suicide-deterring contents such as warning signs or prevention resources. To compare frequency rates of TPB variables, messages that are likely to stimulate perceived behavior control (78.11%) were the most prevalent, followed by subjective norms (15.87%) and attitude (6.02%).

This study suggests an alternative way to classify large-scale text dataset based on a typicality measurement. We first based our approach on LDA to investigate how the semantic space is partitioned by theoretical concepts. In an attempt to conduct LDA based on specific theory, we included a little portion of human-annotated tweets in LDA process, then conducted Nearest Neighbor and typicality-based clustering. This process requires human judgment to determine to which extent typicality value can be accepted as corresponding to each TPB variable. Overall, our model quite accurately classified tweets according to the TPB framework. Taking account of diverse usages of unstandardized words, short-length text, lots of metaphors and sophisticated lexical patterns that capture persuasive features, we were convinced that average classification accuracy of 74.77% is acceptable. As typicality is a simple but powerful technique (Bappy et al., 2017), we were able to minimize demand for train set to learn a classification model. Thus, it may be helpful to future automated content analyses, especially to other research subject with lack of dictionaries or lexical resources for train data.

The contributions of this work are twofold. First of all, this study employed a novel automated text analytic process designed to take advantage of unsupervised learning and supervised learning. Supervised approach is often considered to be appropriate for theory-based content analysis (Grimmer & Stewart, 2013), while manually annotating a large number of train set is a time-consuming task. On the other hand, unsupervised approach requires less human effort but barely yields result that correspond to variables of the specific theory (Schwartz & Ungar, 2015). Although the combination of supervised and unsupervised approach may seem unusual, some studies in text mining (Ramage, Manning, & Dumais, 2011; Shutova et al., 2017) and computer vision (Liu, Rosenberg & Rowley, 2007) have reported the increased classification performance when two distinct approaches are combined. We also took advantage of those two approaches so that we could analyze large-scale data with less human effort.

Secondly, our model classified tweets according to TPB categories: attitude, subjective norm, and PBC. Although this study did not directly test the TPB, it provided a methodological framework for computer to classify messages as similar as the way communication researchers do. This was possible since we focused on how to decide "typicality" of each TPB variable. We showed how detection performance changes as typicality rate alters from 0 to 1. In this process, the rate "partial match with typical pattern" was judged by researchers because human is considered as best at interpreting latent meanings of the message (Krippendorff, 2012). Even though computational approaches are generally known to be inadequate to grasp latent meanings as a human, applying typicality measurement to automated methods may provide more chances to get closer to human's way of interpreting messages. This shows one way to combine benefits of computational tools and human judgment when identifying persuasive contexts in large-scale data.

**Limitations**

While our approach has advantages on classifying large-scale suicide tweets into theoretical variables, it is not without limitations. First of all, our automated model could not detect one category: *approval of significant others*. One possible reason is, tweets containing *approval of significant others* is indeed rare in Twitter. Another reason may lie in inadequate number of human-annotated tweets to identify the typical word distributions of that topic. If more human-annotated tweets are included, then we expect to detect more tweets representing *approval of significant others*.

Second limitation is unsatisfactory detection rates of some categories such as *difficulty of suicide*, *negative outcomes* and *disapproval of others*. We qualitatively analyzed the reasons for low detection accuracy and found out that TPB variables are not limited to just one sentence. In some cases, one has to take other sentences into account to decide whether the text carries variables of TPB. For instance, "Count three, but if you do *not* calm down, you should kill yourself." and "Count three, but if you do calm down, you should *not* kill yourself" are composed of same words. The first tweet should be classified as *approval of others*, while the second one refers to *disapproval of others*. However, as LDA does not count in word order, the computer could not completely discern different meanings between two tweets.

Lastly, we retrieved suicidal tweets which contain the keyword "suicide". However, one search term is not enough to capture all relevant tweets relevant to the suicide issue (Striker et al., 2006). Some tweets that are related to the suicide issue do not include the word "suicide", rather, they use the term "giving up on life" or "disappear", etc. These search terms should also be included while retrieving tweet data. Therefore, future studies should retrieve tweets containing various words which are relevant to the topic.

**Conclusions and Future Directions**

To date, social media such as Twitter is regarded as a complicated black box with it's potential impact on suicide contagion (Schäfer & Quiring, 2015). Thus, detecting persuasive elements on suicidal tweets from the theoretical perspective is first required in order to address the issue. This research provides a summary of the suicidal tweets that is impossible to obtain manually, and introduces a combined computational approach to detect persuasive elements in large-scale text data. As such, this study represents an important step toward integrating theory-driven and data-driven approach for analyzing "big data" in communication research.

THEORY-DRIVEN AUTOMATED CONTENT ANALYSIS OF SUICIDAL TWEETS     24

THEORY-DRIVEN AUTOMATED CONTENT ANALYSIS OF SUICIDAL TWEETS    25thesis, Seoul National University, Seoul, Korea.

Joffe, H., & Haarhoff, G. (2002). Representations of far-flung illnesses: the case of Ebola in Britain. *Social science & medicine*, *54*(6), 955-969. doi:10.1016/S0277-9536(01)00068-5

Krippendorff, K. (2012). *Content analysis: An introduction to its methodology* (3rd ed.). Thousand Oaks, CA: SAGE.

Lee, J., Lee, W. Y., Hwang, J. S., & Stack, S. J. (2014). To what extent does the reporting behavior of the media regarding a celebrity suicide influence subsequent suicides in South Korea?. *Suicide and life-threatening behavior*, *44*(4), 457-472. doi: 10.1111/sltb.12109

Lee, K., Qadir, A., Hasan, S. A., Datla, V., Prakash, A., Liu, J., & Farri, O. (2017). *Adverse Drug Event Detection in Tweets with Semi-Supervised Convolutional Neural Networks*. In *Proceedings of the 26th International Conference on World Wide Web* (pp. 705-714). Perth, Australia: ACM. doi:10.1145/3038912.3052671.

Liu, T., Rosenberg, C., & Rowley, H. A. (2007). Clustering billions of images with large scale nearest neighbor search. In *Applications of Computer Vision, 2007. WACV'07. IEEE Workshop on* (pp. 28-33). Austin, TX: IEEE. doi: 10.1109/WACV.2007.18.

Luxton, D. D., June, J. D., & Fairall, J. M. (2012). Social media and suicide: a public health perspective. *American journal of public health*, *102*(2), 195-200. doi:10.2105/AJPH.2011.300608

Manganello, J., & Blake, N. (2010). A study of quantitative content analysis of health messages in US media from 1985 to 2005. *Health communication*, *25*(5), 387-396.

THEORY-DRIVEN AUTOMATED CONTENT ANALYSIS OF SUICIDAL TWEETS    29

**Table 1**

Distribution of human-annotated tweets

| Label | Example Tweet | Number (proportion) |
|---|---|---|
| *Attitude* | | |
| Positive outcomes | @USER Suicide is the answer | 498 (14.11%) |
| Negative outcomes | My life is mine! Suicide for hell | 66 (1.87%) |
| *Subjective Norm* | | |
| Approval of significant others | My mom says you'll kill yourself | 37 (1.05%) |
| Disapproval of significant others | Why do you die? Suicide prohibited except me | 63 (1.78%) |
| Approval of others | I would seriously recommend suicide, Suicide! | 616 (17.45%) |
| Disapproval of others | Hami: Do not commit suicide anyway | 545 (15.44%) |
| Descriptive norms | Domestic suicide rate is the highest among OECD countries | 398 (11.27%) |
| *Perceived Behavior Control* | | |
| Ease of suicide | I can commit suicide in 10 seconds! | 55 (1.56%) |
| Difficulty of suicide | I've tried suicide five times, but never succeeded | 165 (4.67%) |
| Mention of specific methods | The easiest and fastest way to suicide: Swiss euthanasia 20,000 dollars | 698 (19.77%) |
| Mention of specific place of suicide | Suicide bomber killed at least 70 people in Pakistan hospital (photo,video) https: // t.co/WQQ7ep5ur | 210 (5.95%) |
| Sources of help against suicide | Middle school suicide prevention and mental health campaign: https://t.co/RogrlWYY8i | 98 (2.78%) |
| Sources that promote suicide | Photo suggesting suicide https://t.co.bnS4arE4zb | 81 (2.29%) |
| Total | | 3530 (100%) |

**Table 2**

Frequent words of each TPB variable extracted from LDA

| Frequent words | Label |
| --- | --- |
| 1.stress, end, I, life, suicide, age, grade, hard, before | Positive outcomes |
| 2.tragedy, family, process, change, homosexual, harass, too, factual | Negative outcomes |
| 3.suicide, MENTION, impulse, immorality, sexual, wicked, read, pill, my, become | Approval of significant others |
| 4.parent, suicide, not, friend, surroundings, fool, alive, one, you | Disapproval of significant others |
| 5.better, die, likely, cause, context, my, suicide, leak, absolutely, debt | Approval of others |
| 6.suicide, not, able, really, high, failed, stop, besides, discarded | Disapproval of others |
| 7.suicide, Korean, last, year, month, people, oneself, MENTION, number | Descriptive norms |
| 8.extent, LOL, Hmm, well, bit, rather, tried, moment | Ease of suicide |
| 9.suicide, survive, now, you, don't, day, willpower, without | Difficulty of suicide |
| 10.suicide, terror, bomb, suspect, male, likelihood, estimate, bride, court, breaking | Mention of specific methods |
| 11.suicide, feasible, river, weather, I, disaster, razor, dawn, do | Mention of specific place of suicide |
| 12.URL, MENTION, upload, only, prevention, GIF, shout | Sources of help against suicide |
| 13.suicide, Naver, MENTION, news, site, horrible, tear, notice, URL | Sources that promote suicide |

**Table 3**

Examples of judgment of TPB variables as typicality score alters from 0 to 1

Attitude (positive outcomes)

| Suicide is the answer | Just suicide is the only answer | How long do I have to be painful? I don't know. I just want to kill myself | Reading is fun, restful, comforting and my little.. Commit suicide | I've run out of painkiller. It deserve to be praised that I did not commit suicide |
|---|---|---|---|---|
| $d = 0$ match | $d = 0.173$ partial match | $d = 0.274$ partial match | $d = 0.292$ mismatch | $d = 0.496$ mismatch |

Subjective Norm (descriptive norms)

| Suicide is the number one cause of death in Korea | Even the famous baseball commentator was discovered dead | Korea ranked the top for suicide rates for the OECD nations | Driven by bullying, a worker killed himself. But the company has no responsibility | Save the world from suicide by renewing and reconstructing civilization |
|---|---|---|---|---|
| $d = 0$ match | $d = 0.179$ partial match | $d = 0.270$ partial match | $d = 0.305$ mismatch | $d = 0.499$ mismatch |

Perceived Behavior Control (mention of specific methods)

| Suicide, suicide, euthanasia, suicide. hahaha | Suicide, suicide, euthanasia, dive into Han river, the pain just continues | When I was 11 years old, I saw something jumping out of the apartment. | I didn't intend to commit suicide. I heard that Mr.Gacha and Ms.Kiye were shot by a gun. | Jews were destroyed by the Roman invasion and they chose a extreme way of suicide |
|---|---|---|---|---|
| $d = 0$ match | $d = 0.173$ partial match | $d = 0.274$ partial match | $d = 0.292$ mismatch | $d = 0.498$ mismatch |

**Table 4**

Frequency of TPB variables in suicide tweets

| Label | Proportion ($n = 214{,}580$) | Accuracy |
|---|---|---|
| *Attitude* | 6.02% | |
|    Positive outcomes | 5.76% | 72.82% |
|    Negative outcomes | 0.26% | 20.21% |
| | | |
| *Subjective Norm* | 15.87% | |
|    Approval of significant others | 0% | 0 |
|    Disapproval of significant others | 1.86% | 61.62% |
|    Approval of others | 0.47% | 81.86% |
|    Disapproval of others | 3.73% | 46.35% |
|    Descriptive norms | 9.81% | 80.29% |
| | | |
| *Perceived Behavior Control* | 78.11% | |
|    Ease of suicide | 14.73% | 74.49% |
|    Difficulty of suicide | 2.37% | 14.14% |
|    Mention of specific methods | 39.89% | 76.43% |
|    Mention of specific place of suicide | 1.64% | 72.66% |
|    Sources of help against suicide | 2.89% | 55.97% |
|    Sources that promote suicide | 16.59% | 85.40% |
| Total | 100% | 74.77 |

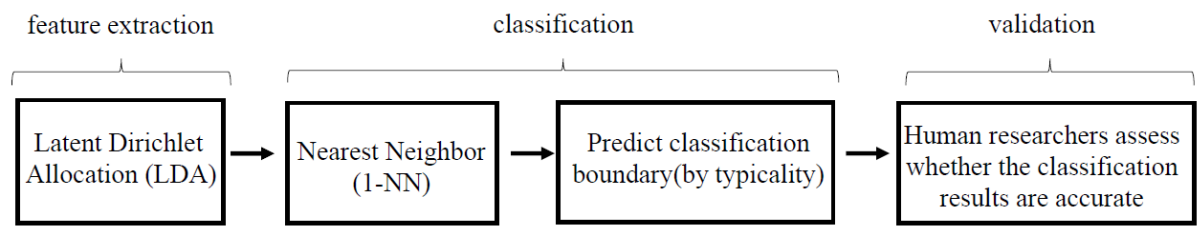

**Figure 1.** Overview of the process

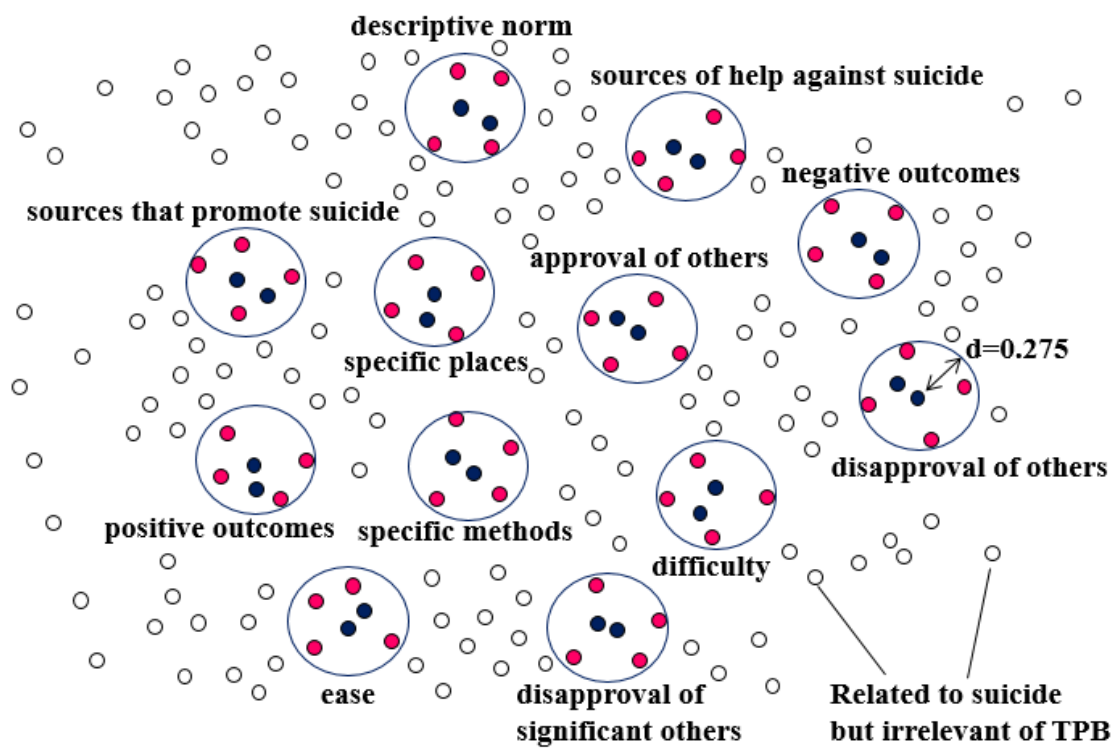

**Figure 2.** Illustration of the idea of TPB-based classification

*Note*: Black points refer to human-annotated tweets. Red points refer to newly labeled tweets which are discerned as "partial match" with human-annotated tweets.